\documentclass[11pt,letterpaper]{article}

\usepackage{amsmath, amssymb}
\usepackage{graphicx}
\usepackage{pstricks}
\usepackage{pst-plot}
\usepackage{pst-math}

\title{Time-Symmetric Action-at-a-Distance Electrodynamics 
and the Principle of Action and Reaction for Particles at Relative Rest}
\author{C\u{a}lin Galeriu \\
Military Technical Academy \lq\lq Ferdinand I\rq\rq \\
Bvd. George Co\c{s}buc nr. 39-49, Bucharest, Romania \\
{\it calin.galeriu@mta.ro}}
\date{}

\begin{document}

\maketitle

\begin{abstract}
{\it We investigate a theory of time-symmetric action-at-a-distance electrodynamic 
or gravitational interactions
where the four-forces depend only on 
the electric charges, the rest masses,
the position four-vectors,
and the four-velocities of the two interacting particles, but not on their 
four-accelerations or higher derivatives \cite{GaleriuArxiv2024}.
The goal is to prove that the principle of action and reaction is obeyed
due to the fact that, for a given pair of corresponding infinitesimal segments
along the worldlines of the two particles, the impulse of the advanced four-force 
and the impulse of the retarded four-force
are equal in magnitude but opposite in direction \cite{Wheeler1949}.
For two particles at relative rest, we derive this result as the outcome of a symmetry operation
in flat conformal space.
We start by assuming a positive spacetime interval between the two interacting particles,
and we perform a conformal inversion (an improper inversion in a hypersphere), after which the two
particles exchange their position and time coordinates and their rest masses.
Then we perform an improper reflection across the axis connecting the two particles,
after which the two particles recover their initial four-velocities, up to a minus sign.
Two improper coordinate transformations are needed together in order to
obtain a resulting positive Jacobian determinant.
In the final step we go to the limit of a null spacetime interval between the 
endpoints of the corresponding infinitesimal segments. 
When the two interacting particles are at rest, or move with the same velocity,
the initial and the final physical parameters that determine the four-forces are identical.
Due to symmetry, the principle of action and reaction emerges.
We make the conjecture that another, undetermined yet, symmetry operation must also apply, in order for the 
principle of action and reaction to hold even for particles moving with different velocities.}
\end{abstract}

\section{Introduction}

The Principle of Action and Reaction (PAR) 
("To every action there is always opposed and equal reaction."
\cite{Newton}), published by Sir Isaac Newton in 1687, has been one of the most
important laws of physics ever since. Together with the impulse-momentum theorem,
$\overrightarrow{F} \, dt = \overrightarrow{d \, p} = \overrightarrow{p_f} - \overrightarrow{p_i}$,
the PAR leads to the law of conservation of total momentum. 
Let $\overrightarrow{F_1}$ be the force with which the second object acts on the first one,
and let $\overrightarrow{F_2}$ be the force with which the first object acts on the second one. Then
\begin{equation}
\begin{array}{l}
\overrightarrow{F_1} \, dt = \overrightarrow{p_{1 f}} - \overrightarrow{p_{1 i}} \\
\overrightarrow{F_2} \, dt = \overrightarrow{p_{2 f}} - \overrightarrow{p_{2 i}} \\
\overrightarrow{F_2} = - \overrightarrow{F_1}
\end{array}
\Bigg\}
\Rightarrow
\overrightarrow{p_{1 i}} + \overrightarrow{p_{2 i}} = \overrightarrow{p_{1 f}} + \overrightarrow{p_{2 f}},
\label{eq:1}
\end{equation}
where $i$ stands for initial and $f$ stands for final.
But notice that this equivalence is true only in Newtonian physics, where we have an absolute time, 
instantaneous interactions,
and the infinitesimal time interval $dt = t_f - t_i$ is the same for both interacting objects.

The law of conservation of total momentum is seen as equivalent
to the PAR, but qualitatively superior, since it is 
directly related to translational symmetry by Noether's theorem. 

When the interaction force is conservative, the action and reaction forces are obtained from a potential function 
$V(\overrightarrow{r_1}, \overrightarrow{r_2})$. 
In this case  $\overrightarrow{F_1} = - \nabla_1 V$ and $\overrightarrow{F_2} = - \nabla_2 V$.
The translational symmetry, expressed by 
$V(\overrightarrow{r_1}, \overrightarrow{r_2}) = V(\overrightarrow{r_1} - \overrightarrow{r_2}, 0)$, 
straightforwardly leads us to the PAR.

But when the generalized potential function 
$V(\overrightarrow{r_1}, \overrightarrow{r_2}, \overrightarrow{v_1}, \overrightarrow{v_2})$ also depends
on the velocities of the interacting particles, like in the Biot-Savart law, 
the action and reaction forces can have different directions,
and the PAR clearly no longer applies \cite{Goldstein2ed}.
This is a failure of Newtonian physics, with its instantaneous action at a distance, 
and not a failure of the PAR {\it per se}.
It is therefore of fundamental importance to investigate what happens to the PAR during the
transition from Newtonian to relativistic physics. 

In relativistic physics, instead of forces $\overrightarrow{F}$ and momenta $\overrightarrow{p}$, 
we have four-forces $\overrightarrow{\bf F}$ and four-momenta $\overrightarrow{\bf p}$.
Instead of (\ref{eq:1}) we now write
\begin{equation}
\begin{array}{l}
\overrightarrow{\bf F_1} \, d\tau_1 = \overrightarrow{\bf p_{1 f}} - \overrightarrow{\bf p_{1 i}} \\
\overrightarrow{\bf F_2} \, d\tau_2 = \overrightarrow{\bf p_{2 f}} - \overrightarrow{\bf p_{2 i}} \\
\overrightarrow{\bf p_{1 i}} + \overrightarrow{\bf p_{2 i}} = \overrightarrow{\bf p_{1 f}} + \overrightarrow{\bf p_{2 f}}
\end{array}
\Bigg\}
\Rightarrow
\overrightarrow{\bf F_2} \, d\tau_2 = - \overrightarrow{\bf F_1} \, d\tau_1,
\label{eq:2}
\end{equation}
where $d\tau_1$ and $d\tau_2$ are some infinitesimal proper time intervals. 
But since in relativistic physics the interactions are no longer instantaneous,
and in order to have a relationship between the same corresponding worldline elements,
in this pair of four-forces, $\overrightarrow{\bf F_1}$ and $\overrightarrow{\bf F_2}$, 
one of them must have a retarded effect and the other one must have an advanced effect.

The changes in the four-momenta of the two objects could be imagined as the result of an interaction mediated by exchange particles.
For example, we can assume that the first object emits an exchange particle with four-momentum 
$\overrightarrow{\bf p} = \overrightarrow{\bf p_{1 i}} - \overrightarrow{\bf p_{1 f}}$, 
which is later absorbed by the second object, whose momentum becomes
$\overrightarrow{\bf p_{2 f}} = \overrightarrow{\bf p_{2 i}} + \overrightarrow{\bf p}$.
While this approach is fully compatible with the law of conservation of total four-momentum, it sends us on the wrong track.
That is because the emission and the absorption of an exchange particle are instantaneous processes, 
and using them hides the relationship between the infinitesimal proper time intervals $d\tau_1$ and $d\tau_2$.

Another approach that hides the relationship between $d\tau_1$ and $d\tau_2$ 
was explored by Houtappel, Van Dam, and Wigner
\cite{Houtappel1965}, who considered an interaction that extends from a retarded to an advanced time on the worldline of the 
particle that produces the four-force
\begin{equation}
\overrightarrow{{\bf F_2}}(\tau_2) = \int \overrightarrow{{\bf F_{21}}}(\tau_2, \tau_1) \, d\tau_1,
\label{eq:HVDW_1}
\end{equation}
\begin{equation}
\overrightarrow{{\bf F_1}}(\tau_1) = \int \overrightarrow{{\bf F_{12}}}(\tau_1, \tau_2) \, d\tau_2,
\label{eq:HVDW_2}
\end{equation}
and write the PAR in terms of these four-force linear densities
\begin{equation}
\overrightarrow{{\bf F_{21}}}(\tau_2, \tau_1) = - \overrightarrow{{\bf F_{12}}}(\tau_1, \tau_2).
\label{eq:HVDW_3}
\end{equation}

In order to explicitly exhibit the relationship between $d\tau_1$ and $d\tau_2$,
it is much better to imagine the interaction as taking place between 
corresponding infinitesimal segments on the worldlines of the two objects.
By talking about worldlines (instead of worldtubes) we are in fact implying that the two objects are very small, 
and that in 3D space the material point model fits them well. In practice we are dealing with electrically charged 3D point particles.
In this interaction model the corresponding infinitesimal segments, of length $ds_1 = i\,c\,d\tau_1$ and $ds_2 = i\,c\,d\tau_2$, 
have their end points connected by light signals \cite{fokker}.
This relationship is not affected by either Lorentz or conformal transformations.
In fact, the Lorenz group, which leaves the length of any spacetime intervals invariant, 
is a subgroup of the conformal group, which leaves the length of only null spacetime intervals invariant.

It was known for a long time that Maxwell's equations are invariant
not only under the Lorentz group, but also under the
conformal group \cite{bateman1, cunningham, bateman2}, 
and that the equations of motion are invariant as well, 
provided that the rest mass transforms like a 
reciprocal length \cite{PageAdams1936, SchoutenHaantjes1936, Haantjes1940, Bopp1959}.

A strong hint that the conformal group might play an essential role is brought by Zeeman's theorem \cite{Zeeman}.
If causality implies the Lorentz group, what happens when we give up causality and we consider a time-symmetric interaction?
We are then no longer restricted to the symmetry of the inhomogeneous Lorentz group and dilatations. 
The most natural extension of the Lorentz group brings to our attention the full conformal group.
However, historically, due to the puzzling variation of the rest mass, the favourable reception
of conformal invariance in classical physics has been avoided \cite{Rohrlich2007}.

While we readily accept the fact that the emission or the absorption of a real photon by an atom changes the rest mass of this atom, 
we don't believe that the same thing happens when an electrically charged elementary particle
emits or absorbs a virtual photon. This quantum electrodynamics dictum is inherited from classical electrodynamics.
The variation of the rest mass is something that simply doesn't happen in classical relativistic physics, due to the
antisymmetry of the field strength tensor. This antisymmetry is the result of the assumption that the 
electromagnetic field is fully described by a potential four-vector field, and that the interaction term
in the Lagrangian is proportional to the scalar product of the four-potential of the field with the four-velocity of the 
electrically charged particle. Could it be that the classical theory has made us believe only 
an oversimplified description, that the electromagnetic field might exhibit a more complex structure,
and that the rest mass of elementary particles might vary? 
Geometrical arguments in Minkowski space have led us to exactly this conclusion \cite{GaleriuArxiv2024}.
A variable rest mass could be accommodated as a scalar field in Minkowski space \cite{Galeriu2004},
or as the fifth coordinate of the position vector in a 5D space-time-mass manifold, 
or as the fifth component of the momentum-energy-mass vector \cite{Leclerc2002}.

Since in Newtonian physics the PAR is equivalent to the law of conservation of total momentum, which can be directly derived
from translational symmetry considerations, it seems natural to ask whether in relativistic physics the PAR could also
be directly derived from some symmetry considerations. However, as it will soon be quite clear in Section 2,
in order to give a positive answer to the  
previously stated question, in the time-symmetric theoretical framework considered,
we will have to find a coordinate transformation that can swap 
the position, the time, the velocity, and the rest mass of the two 
interacting particles. 
Due to the required variation of the rest mass, 
and based on the arguments presented so far, 
it seems natural to suspect that conformal symmetry is responsible for the PAR.

In Section 2 we derive the PAR as the outcome of a symmetry operation in Minkowski space,
but only for two identical elementary particles (with the same rest mass) at relative rest.
In Section 3 we review important formulas related to conformal inversions.
In Section 5 we review important formulas related to reflections across an axis.
In Sections 4, 6, and 7 we derive the PAR as the outcome of a symmetry operation in flat conformal space,
but only for two different elementary particles (with different rest masses) at relative rest.

\section{A derivation of the PAR for two identical elementary particles at relative rest}

Consider two identical elementary particles in Minkowski space.
We assume that both particles are at rest, or moving with the same velocity, in which
case we bring them to rest by performing a Lorentz transformation.
The two interacting point particles are connected by a spacetime interval of null length.
It is clear that in these circumstances any two corresponding worldline segments have the same 
length, which means that $d\tau_1 = d\tau_2$, and the PAR simplifies to 
$\overrightarrow{\bf F_2} = - \overrightarrow{\bf F_1}$.

Let particle $A$ have the position four-vector $ (x_A, y_A, z_A, i\,c\,t_A)$
and let particle $B$ have the position four-vector $(x_B, y_B, z_B, i\,c\,t_B)$.
We assume that $t_A < t_B$.

We investigate a theory of time-symmetric action-at-a-distance interactions
where particle $A$ exerts 
a retarded four-force $\overrightarrow{\bf F_B}$ on particle $B$,
and where particle $B$ exerts 
an advanced four-force $\overrightarrow{\bf F_A}$ on particle $A$.
For the derivation of the PAR we focus only on this pair of four-forces,
$\overrightarrow{\bf F_A}$ and $\overrightarrow{\bf F_B}$,
although in our time-symmetric theory 
the total four-force acting on a particle at any given spacetime point is in general a sum
of two contributions, one retarded and one advanced. 

Under a time reversal operation 
\footnote{In the active view of time reversal the velocity and the magnetic field vectors change direction
($\vec{v} \to - \vec{v}$, $\vec{B} \to - \vec{B}$) while the electric field and the force vectors don't
($\vec{E} \to \vec{E}$, $\vec{F} \to \vec{F}$). 
Only the spatial components of the four-velocity change sign.
Only the temporal component of the four-force changes sign.
In the passive view of time reversal the new time axis points into the opposite direction ($t' = - t$)
and the four-velocity vectors change direction ($\overrightarrow{\bf U'} = - \overrightarrow{\bf U}$).
The four-force components are left unchanged, since
the minus sign brought by the change in the
direction of the four-momenta
is cancelled by the minus sign brought by
the swap of the initial and final four-momenta.}
the four-force $\overrightarrow{\bf F_B}$ 
that was retarded becomes advanced,
and the four-force $\overrightarrow{\bf F_A}$ that was advanced becomes retarded.
But, from a geometrical point of view, 
the Minkowski diagram looks the same.
The total four-force, being the sum of the same two contributions, 
has the same magnitude and 
makes the same angle with the tangent to the worldline of the particle on
whom it acts. 
By symmetry, the motion proceeds in reverse, with the 3D velocities of
the particles pointing into the opposite direction.

Let $O$ be the midpoint of segment $AB$. We translate the origin of our 
4D reference frame to point $O$. As a consequence, we are now able to write
\begin{eqnarray}
x_B = - x_A, \\
y_B = - y_A, \\
z_B = - z_A, \\
i\,c\,t_B = - i\,c\,t_A.
\end{eqnarray}

We perform a spacetime inversion through the origin $O$, which is the well known 
parity and time reversal (PT) operation. For two elementary particles with the same rest mass
but electric charges of opposite sign, we also need a charge conjugation (C) operation, but
for now we ignore this possibility that brings nothing new to the discussion.
As a result of the PT operation the new spacetime coordinates become
\begin{eqnarray}
x' = - x, \\
y' = - y, \\
z' = - z, \\
i\,c\,t' = - i\,c\,t.
\end{eqnarray}
The Jacobian matrix of this coordinate transformation is
\begin{equation}
 \left( \frac{\partial x^\mu}{\partial x'^\nu} \right) = 
\begin{pmatrix}
-1 & 0 & 0 & 0 \\ 0 & -1 & 0 & 0 \\ 0 & 0 & -1 & 0 \\ 0 & 0 & 0 & -1
\end{pmatrix},
\label{JacobianPT}
\end{equation}
where $x^1 = x$, $x^2 = y$, $x^3 = z$, and $x^4 = i\,c\,t$.
The determinant of this Jacobian matrix is 1.

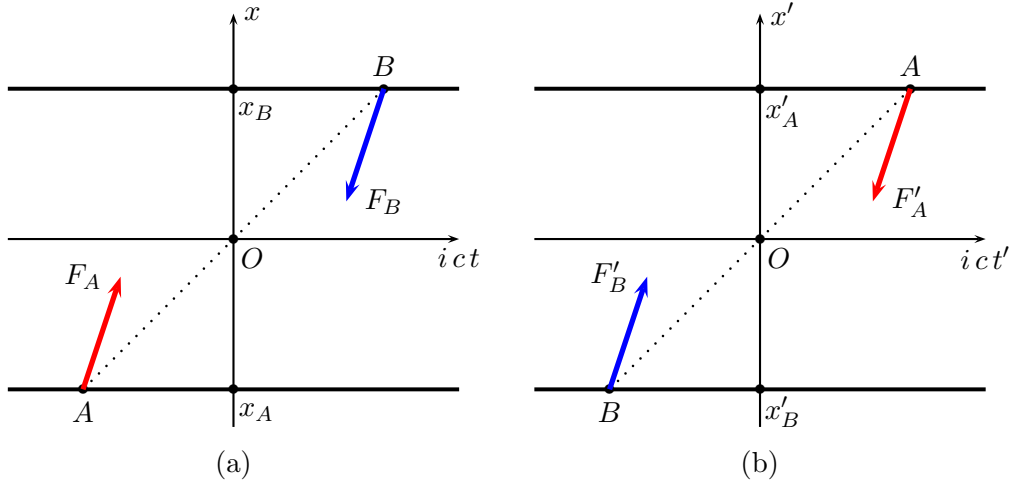
\begin{figure}[h!]
\begin{center}
\begin{pspicture}(-3,-3)(10,3)
% left side
\psdot(0,0)
\rput(0.25,-0.25){$O$}
\psline{->}(-3,0)(3,0)
\rput(3,-0.25){$i\,c\,t$}
\psline{->}(0,-2.5)(0,3)
\rput(0.25,3){$x$}
\psline[linewidth=1.5pt](-3,-2)(3,-2)
\psline[linewidth=1.5pt](-3,2)(3,2)
\psdot(-2,-2)
\rput(-2,-2.3){$A$}
\psdot(2,2)
\rput(2,2.3){$B$}
\psdot(0,-2)
\rput(0.3,-2.3){$x_A$}
\psdot(0,2)
\rput(0.3,1.7){$x_B$}
\psline[linestyle=dotted,linewidth=1pt](-2,-2)(2,2)
\psline[linewidth=2pt,linecolor=red]{->}(-2,-2)(-1.5,-0.5)
\rput(-2,-0.5){$F_A$}
\psline[linewidth=2pt,linecolor=blue]{->}(2,2)(1.5,0.5)
\rput(2,0.5){$F_B$}
\rput(0,-3){(a)}
% right side
\psdot(7,0)
\rput(7.25,-0.25){$O$}
\psline{->}(4,0)(10,0)
\rput(10,-0.25){$i\,c\,t'$}
\psline{->}(7,-2.5)(7,3)
\rput(7.3,3){$x'$}
\psline[linewidth=1.5pt](4,-2)(10,-2)
\psline[linewidth=1.5pt](4,2)(10,2)
\psdot(5,-2)
\rput(5,-2.3){$B$}
\psdot(9,2)
\rput(9,2.3){$A$}
\psdot(7,-2)
\rput(7.3,-2.3){$x'_B$}
\psdot(7,2)
\rput(7.3,1.7){$x'_A$}
\psline[linestyle=dotted,linewidth=1pt](5,-2)(9,2)
\psline[linewidth=2pt,linecolor=blue]{->}(5,-2)(5.5,-0.5)
\rput(5,-0.5){$F'_B$}
\psline[linewidth=2pt,linecolor=red]{->}(9,2)(8.5,0.5)
\rput(9,0.5){$F'_A$}
\rput(7,-3){(b)}
\end{pspicture}
\caption{The spacetime configurations before and after the PT operation.}
\label{fig:1}
\end{center}
\end{figure}

As a result of the PT operation, 
particle $A$ acquires the initial coordinates of particle $B$,
\begin{eqnarray}
x'_A = - x_A = x_B, \\
y'_A = - y_A = y_B, \\
z'_A = - z_A = z_B, \\
i\,c\,t'_A = - i\,c\,t'_A = i\,c\,t_B,
\end{eqnarray}
and particle $B$ acquires the initial coordinates of particle $A$,
\begin{eqnarray}
x'_B = - x_B = x_A, \\
y'_B = - y_B = y_A, \\
z'_B = - z_B = z_A, \\
i\,c\,t'_B = - i\,c\,t'_B = i\,c\,t_A.
\end{eqnarray}
In fact, as seen in Figure \ref{fig:1}, the whole worldlines of the two particles at rest exchange
their places in Minkowski space, in the passive view of this PT symmetry operation.

As a result of the PT operation, the four-force acting on particle $A$
changes direction
\begin{equation}
{F'_A}^\mu = \frac{\partial x'^\mu}{\partial x^\nu} F_A^\nu = - \delta^\mu_\nu \, F_A^\nu = - F_A^\mu, 
\label{eq:2_23}
\end{equation}
and the same thing happens to the four-force acting on particle $B$
\begin{equation}
{F'_B}^\mu = \frac{\partial x'^\mu}{\partial x^\nu} F_B^\nu = - \delta^\mu_\nu \, F_B^\nu = - F_B^\mu. 
\label{eq:2_24}
\end{equation}

Since the two elementary particles at rest have the same rest mass and
electric charge, when their worldlines exchange positions in Minkowski space 
due to the PT operation, we
obtain a final spacetime configuration undistinguishable from 
the initial spacetime configuration.
As a consequence, due to symmetry, the following two relations must hold
\begin{eqnarray}
{F'_A}^\mu = F_B^\mu, \label{eq:2_25} \\
{F'_B}^\mu = F_A^\mu. \label{eq:2_26}
\end{eqnarray}

Due to equations (\ref{eq:2_23})-(\ref{eq:2_24}), 
either of equations (\ref{eq:2_25})-(\ref{eq:2_26}) is equivalent to
\begin{equation}
\overrightarrow{\bf F_B} = - \overrightarrow{\bf F_A},
\end{equation}
and in this way the PAR emerges as a consequence of PT (or CPT) invariance.

When looking closer at this derivation of the PAR, it comes to light the fact that,
as a result of the PT operation, non only the four-forces change sign, but also
the four-velocities, like any of the involved four-vectors.
The four-velocity of particle $A$ changes direction
\begin{equation}
{U'_A}^\mu = \frac{\partial x'^\mu}{\partial x^\nu} U_A^\nu = - \delta^\mu_\nu \, U_A^\nu = - U_A^\mu, 
\label{eq:2_28}
\end{equation}
and the same thing happens to the four-velocity of particle $B$
\begin{equation}
{U'_B}^\mu = \frac{\partial x'^\mu}{\partial x^\nu} U_B^\nu = - \delta^\mu_\nu \, U_B^\nu = - U_B^\mu. 
\label{eq:2_29}
\end{equation}

This is of no concern to us. In Minkowski space nothing moves anyway. The past, the present, and the future 
\lq\lq are all {\it equally} existent\rq\rq \cite{VesselinPetkov}.
The worldlines, the tangent lines to the worldlines, and the curvatures of the
worldlines are pure geometric objects that describe, 
in a time symmetric invariant way, the dynamics of the interacting particles.
\lq\lq It is a basic fact of electromagnetic life that the net electromagnetic force
experienced by the particle at a point depends only on its tangent line $L$ there and its
charge $q$.\rq\rq\ \cite{Malament2004}
The positive or negative sign of the four-velocity is irrelevant, it is nothing but a consequence of 
our freedom of choice when, in Minkowski space, we introduce a continuous timelike vector field $\tau^a$
that determines a temporal orientation. 
\lq\lq Rather than representing the particle’s instantaneous velocity at
a point as a tangent line, we represent it as a unit timelike vector there (co-aligned
with the tangent line). But there are two from which to choose. One is
future-directed with respect to $\tau^a$; the other is past-directed. It makes no difference
which we choose [...].\rq\rq\ \cite{Malament2004}

What happens when the two elementary particles have different rest masses?
For a similar symmetry argument to work, we would need a coordinate transformation that
also brings an exchange in the values of the rest masses of the two interacting particles.
For this reason we focus our attention on conformal transformations, 
and in particular on conformal inversion, which is 
\lq\lq one of the sharpest tools available to the conformal theorist\rq\rq \cite{NicholasWheeler}. 

\section{A review of conformal inversion formulas}

Consider the conformal inversion transformation in flat spacetime
\begin{equation}
x'^\mu = L^2 \frac{x^\mu}{x^2},
\label{eq:4}
\end{equation}
where $x^2 = g_{\alpha \beta} \, x^\alpha \, x^\beta$,
$g_{\alpha \beta}$ is the metric tensor,
and $L$, the radius of the hypersphere, is an arbitrary constant with units of length.
The inverse transformation looks just like the original transformation, 
the conformal inversion operation being the inverse of itself
\begin{equation}
x^\mu = L^2 \frac{x'^\mu}{x'^2},
\label{eq:5}
\end{equation}
{\it Proof.} Using (\ref{eq:4}) we calculate $x'^2$
\begin{equation}
x'^2 = g_{\alpha \beta} \, x'^\alpha \, x'^\beta 
= g_{\alpha \beta} \, L^2 \frac{x^\alpha}{x^2}  L^2 \frac{x^\beta}{x^2}
= \frac{L^4}{x^2},
\label{eq:6x2}
\end{equation}
which is then substituted into the right side of (\ref{eq:5}), together with (\ref{eq:4}).

Under the conformal inversion transformation the coordinates of a second point transform according to
\begin{equation}
y'^\mu = L^2 \frac{y^\mu}{y^2},
\label{eq:7}
\end{equation}
and we also have
\begin{equation}
y'^2 = g_{\alpha \beta} \, y'^\alpha \, y'^\beta 
= g_{\alpha \beta} \, L^2 \frac{y^\alpha}{y^2}  L^2 \frac{y^\beta}{y^2}
= \frac{L^4}{y^2},
\label{eq:6y2}
\end{equation}
and
\begin{equation}
x' \cdot y' = g_{\alpha \beta} \, x'^\alpha \, y'^\beta 
= g_{\alpha \beta} \, L^2 \frac{x^\alpha}{x^2} L^2 \frac{y^\beta}{y^2}
= L^4 \frac{x \cdot y}{x^2 \, y^2}.
\label{eq:6xy}
\end{equation}
The distance between the two points 
\begin{equation}
(x - y)^2 =  x^2 - 2 x \cdot y + y^2,
\label{eq:dist}
\end{equation}
transforms according to
\begin{equation}
(x' - y')^2 = x'^2 - 2 x' \cdot y' + y'^2
= \frac{L^4}{x^2} - 2 L^4 \frac{x \cdot y}{x^2 \, y^2}+ \frac{L^4}{y^2} = L^4 \frac{(x - y)^2}{x^2 \, y^2}.
\label{eq:8}
\end{equation}

Equation (\ref{eq:8}) shows that, under conformal inversion, 
a light cone transforms into a light cone.
Any two space-time points connected by a light signal, for which $(x - y)^2 = 0$,  
will remain connected by a light signal, because $(x' - y')^2 = 0$ as well after the 
conformal inversion transformation.

Finally, in the limiting case when $x$ and $y$ are separated by an infinitesimal distance, 
$(x - y)^2 = (ds)^2$ and $(x' - y')^2 = (ds')^2$, and we obtain
\begin{equation}
(ds')^2 = \frac{L^4}{x^4} (ds)^2,
\label{eq:10}
\end{equation}
from which, assuming that $L^2 / x^2 > 0$, we obtain
\begin{equation}
ds' = \frac{L^2}{x^2} ds,
\label{eq:11}
\end{equation}
a formula which shows 
that only infinitesimal null line elements are conserved by conformal inversion transformations.
In this first part of the derivation we have closely followed the similar calculations made by Ryder 
for special conformal transformations \cite{Ryder1974, Galeriu_RBEF}.
Although the non-null length of infinitesimal segments is not conserved, ratios of such segments at the same space-time point 
stay the same, which means that (locally) angles are conserved. This is the origin of the 
\lq\lq conformal\rq\rq\ name given to these transformations. 

When calculating the scalar products in 
(\ref{eq:6x2}), (\ref{eq:6y2}), and (\ref{eq:6xy}) we have used 
the same metric tensor $g_{\alpha \beta}$, 
essentially making the hidden assumption that $g'_{\alpha \beta} = g_{\alpha \beta}$.
However, if we look at (\ref{eq:4}) as a coordinate transformation within the framework of General Relativity, we will 
get a different metric tensor. In General Relativity the metric tensor transforms as
\begin{equation}
g'^{(GR)}_{\mu \nu} = \frac{\partial x^\alpha}{\partial x'^\mu} \frac{\partial x^\beta}{\partial x'^\nu} g_{\alpha \beta}.
\label{eq:metricGR}
\end{equation}
The infinitesimal displacements always transform as contravariant vectors
\begin{equation}
dx'^\mu = \frac{\partial x'^\mu}{\partial x^\sigma} dx^\sigma,
\label{eq:dxmuGR}
\end{equation}
\begin{equation}
dx'^\nu = \frac{\partial x'^\nu}{\partial x^\rho} dx^\rho.
\label{eq:dxnuGR}
\end{equation}
From (\ref{eq:metricGR}) - (\ref{eq:dxnuGR}) it follows that the length of the line element is invariant
\begin{multline}
(ds'^{(GR)})^2 = g'^{(GR)}_{\mu \nu} \, dx'^\mu \, dx'^\nu 
= \frac{\partial x^\alpha}{\partial x'^\mu} \frac{\partial x^\beta}{\partial x'^\nu} g_{\alpha \beta}
\frac{\partial x'^\mu}{\partial x^\sigma} dx^\sigma \frac{\partial x'^\nu}{\partial x^\rho} dx^\rho \\
= \frac{\partial x^\alpha}{\partial x^\sigma} \frac{\partial x^\beta}{\partial x^\rho} 
g_{\alpha \beta} \, dx^\sigma \, dx^\rho
= \delta^\alpha_{\ \sigma} \, \delta^\beta_{\ \rho} \, g_{\alpha \beta} \, dx^\sigma \, dx^\rho
= g_{\sigma \rho} \, dx^\sigma \, dx^\rho = (ds)^2,
\label{eq:dsGR}
\end{multline}
a well known result that is very different from what we have in (\ref{eq:10}). 

In order to understand this apparent paradox, using (\ref{eq:5}) we calculate
\begin{multline}
\frac{\partial x^\alpha}{\partial x'^\mu} 
= \frac{\partial}{\partial x'^\mu} \left( \frac{L^2 \, x'^\alpha}{g_{\sigma \rho} \, x'^\sigma \, x'^\rho} \right) \\
= \frac{L^2}{g_{\sigma \rho} \, x'^\sigma \, x'^\rho} \frac{\partial x'^\alpha}{\partial x'^\mu} 
- \frac{L^2 \, x'^\alpha}{( g_{\sigma \rho} \, x'^\sigma \, x'^\rho )^2}
\left( g_{\sigma \rho} \frac{\partial x'^\sigma}{\partial x'^\mu} x'^\rho 
+ g_{\sigma \rho} \, x'^\sigma \frac{\partial x'^\rho}{\partial x'^\mu} \right) \\
= \frac{L^2}{x'^2} \delta^\alpha_{\ \mu} 
- \frac{L^2 \, x'^\alpha}{( x'^2 )^2}
\left( g_{\sigma \rho} \, \delta^\sigma_{\ \mu} \, x'^\rho 
+ g_{\sigma \rho} \, x'^\sigma \, \delta^\rho_{\ \mu} \right) \\
= \frac{L^2}{x'^2} \delta^\alpha_{\ \mu} 
- \frac{L^2 \, x'^\alpha}{( x'^2 )^2}
\left( \delta^\sigma_{\ \mu} \, x'_\sigma + x'_\rho \, \delta^\rho_{\ \mu} \right)
= \frac{L^2}{x'^2} \delta^\alpha_{\ \mu} 
- \frac{2 L^2}{( x'^2 )^2} x'^\alpha \, x'_\mu,
\label{eq:dxadxpm}
\end{multline}
\begin{equation}
\frac{\partial x^\beta}{\partial x'^\nu} 
= \frac{L^2}{x'^2} \delta^\beta_{\ \nu} 
- \frac{2 L^2}{( x'^2 )^2} x'^\beta \, x'_\nu.
\label{eq:dxbdxpn}
\end{equation}
Direct substitution of (\ref{eq:dxadxpm}) and (\ref{eq:dxbdxpn}) into (\ref{eq:metricGR}) 
produces an expression
\begin{multline}
g'^{(GR)}_{\mu \nu}
= \left( \frac{L^2}{x'^2} \delta^\alpha_{\ \mu} - \frac{2 L^2}{( x'^2 )^2} x'^\alpha \, x'_\mu \right) 
\left( \frac{L^2}{x'^2} \delta^\beta_{\ \nu} - \frac{2 L^2}{( x'^2 )^2} x'^\beta \, x'_\nu \right) g_{\alpha \beta} \\
= \frac{L^4}{(x'^2)^2} \delta^\alpha_{\ \mu} \, \delta^\beta_{\ \nu} \, g_{\alpha \beta}
- \frac{2 L^4}{( x'^2 )^3} \delta^\alpha_{\ \mu} \, x'^\beta \, x'_\nu \, g_{\alpha \beta} \\
- \frac{2 L^4}{( x'^2 )^3} x'^\alpha \, x'_\mu \, \delta^\beta_{\ \nu} \, g_{\alpha \beta}
+ \frac{4 L^4}{( x'^2 )^4} x'^\alpha \, x'_\mu \, x'^\beta \, x'_\nu \, g_{\alpha \beta} \\
= \frac{L^4}{(x'^2)^2} g_{\mu \nu}
- \frac{2 L^4}{( x'^2 )^3} x'^\beta \, x'_\nu \, g_{\mu \beta}
- \frac{2 L^4}{( x'^2 )^3} x'^\alpha \, x'_\mu \, g_{\alpha \nu}
+ \frac{4 L^4}{( x'^2 )^4} x'_\mu \, x'_\nu \, x'^\alpha \, x'^\beta \, g_{\alpha \beta},
\label{eq:metricGRvalue}
\end{multline}
which clearly shows that the $g'_{\alpha \beta} = g_{\alpha \beta}$ assumption does not hold.
If we assume that $g'_{\alpha \beta} = g_{\alpha \beta}$ then the last three
terms cancel out (due to the fact that $x'^\beta \, g'_{\mu \beta} = x'_\mu$, and that $x'^\alpha \, g'_{\alpha \nu} = x'_\nu$, 
and also that $x'^\alpha \, x'^\beta \, g'_{\alpha \beta} = x'^2$) and we are left with a contradiction.

In order to save the $g'_{\alpha \beta} = g_{\alpha \beta}$ assumption, 
we have to replace the transformation law (\ref{eq:metricGR}) with
\begin{equation}
g'_{\mu \nu} = \frac{(x'^2)^2}{L^4}
\frac{\partial x^\alpha}{\partial x'^\mu} \frac{\partial x^\beta}{\partial x'^\nu} g_{\alpha \beta},
\label{eq:metricCR}
\end{equation}
which means that we have to replace the pseudo-Riemannian space of General Relativity 
with the flat Weyl space of Conformal Relativity. 

In this Weyl space a tensor density of weight $N$ transforms according to
\begin{equation}
T'^{\mu}_{\ \ \nu} = W^N \, \frac{\partial x'^\mu}{\partial x^\alpha} 
\frac{\partial x^\beta}{\partial x'^\nu} T^{\alpha}_{\ \beta},
\label{eq:tensor_density}
\end{equation}
where $W$ is the Jacobian determinant of the coordinate transformation from $x$ to $x'$. 
Scalars, vectors, and higher rank tensors have zero weight.
A table with the weights of all the important tensor densities that
appear in classical electrodynamics is provided in Ref. \cite{Galeriu_RBEF}.
One may notice that many of these weights are fractional numbers,
which seems to imply the fact that the Jacobian determinant $W$ of any coordinate transformation
must be a positive number. This assumption 
(\lq\lq We limit ourselves to gauge transformations with positive $\lambda$.\rq\rq)
is stated in a footnote by Synge and Schild \cite{SyngeSchild}.

The Jacobian determinant $W$ of transformation (\ref{eq:4}) is \cite{NicholasWheeler}
\begin{equation}
W = \det \left( \frac{\partial x^\mu}{\partial x'^\nu} \right)
= - \frac{L^8}{(x'^2)^4}
= - \left( \frac{L^2}{x'^2} \right) ^4
= - \left( \frac{x^2}{L^2} \right) ^4.
\label{eq:Jacobian}
\end{equation}
The expression (\ref{eq:Jacobian}) of the Jacobian determinant can be obtained 
by factoring out $- L^2 / x'^2$ from each of the matrix elements (\ref{eq:dxadxpm}). 
The determinant of the matrix that is left is equal to $-1$, as demonstrated in Section 5.
The negative value of this Jacobian determinant is troublesome. For this reason
one conformal inversion transformation must always be coupled to
another conformal inversion transformation, as it actually happens in 
special conformal transformations \cite{NicholasWheeler}, or to another improper coordinate
transformation with negative Jacobian determinant. 
Whenever we combine two or more coordinate transformations, 
due to the \lq\lq transitivity of the Jacobian\rq\rq\ property \cite{SyngeSchild, Eddington},
we also multiply their Jacobians. In this way the resulting Jacobian determinant is positive.

Now we can write (\ref{eq:metricCR}) as 
\begin{equation}
g'_{\mu \nu} = |W|^{- \frac{1}{2}} \, 
\frac{\partial x^\alpha}{\partial x'^\mu} \frac{\partial x^\beta}{\partial x'^\nu} g_{\alpha \beta},
\label{eq:metricCRweight}
\end{equation}
which shows that in Conformal Relativity the covariant metric tensor density $g_{\alpha \beta}$ has a weight of $-1/2$. 
The absolute value bars in (\ref{eq:metricCRweight}) 
remind us of the need to couple every
conformal inversion transformation to another improper coordinate transformation with
negative Jacobian determinant.
In this second part of the derivation we have closely followed the similar calculations made by Barut and Haugen \cite{BarutHaugen, Galeriu_RBEF} 
for special conformal transformations.

Suppose that we start with a covariant tensor component $S_\mu$. 
If we raise an index, $S^\alpha = S_\mu \, g^{\mu \alpha}$, 
and then we lower the same index, $S_\nu = S^\alpha \, g_{\alpha \nu}$, 
we get back the same tensor component.
As a result
\begin{equation}
g^{\mu \alpha} \, g_{\alpha \nu} = \delta^{\mu}_{\ \nu}.
\label{eq:Kronecker}
\end{equation}
Since the Kronecker delta $\delta^{\mu}_{\ \nu}$ is a universal tensor of weight 0 
\begin{equation}
\delta'^{\mu}_{\ \nu} 
= \frac{\partial x'^\mu}{\partial x^\alpha} \frac{\partial x^\beta}{\partial x'^\nu} 
\delta^{\alpha}_{\ \beta} 
= \frac{\partial x'^\mu}{\partial x^\alpha} \frac{\partial x^\alpha}{\partial x'^\nu} 
= \frac{\partial x'^\mu}{\partial x'^\nu}
= \delta^{\mu}_{\ \nu},
\label{eq:Kronecker2}
\end{equation}
and since we need to have the same total weight on the left side of the tensorial equation (\ref{eq:Kronecker}),
the contravariant metric tensor density $g^{\alpha \beta}$ must have a weight of 1/2
\begin{equation}
g'^{\mu \nu} = |W|^{\frac{1}{2}} \, 
\frac{\partial x'^\mu}{\partial x^\alpha} \frac{\partial x'^\nu}{\partial x^\beta} g^{\alpha \beta},
\label{eq:metricCRweight2}
\end{equation}
a result consistent with $g'^{\mu \nu} = g^{\mu \nu}$, as can be checked by direct
substitution of (\ref{eq:Jacobian}) and
\begin{equation}
\frac{\partial x'^\mu}{\partial x^\alpha} 
= \frac{L^2}{x^2} \delta^\mu_{\ \alpha} - \frac{2 L^2}{( x^2 )^2} x^\mu \, x_\alpha,
\label{eq:dxpmdxa}
\end{equation}
\begin{equation}
\frac{\partial x'^\nu}{\partial x^\beta} 
= \frac{L^2}{x^2} \delta^\nu_{\ \beta} - \frac{2 L^2}{( x^2 )^2} x^\nu \, x_\beta,
\label{eq:dxpndxb}
\end{equation}
into (\ref{eq:metricCRweight2}). In flat conformal space the metric tensor $g_{\mu \nu}$
is simply the Minkowski metric tensor $\eta_{\mu \nu}$.

\section{A derivation of the PAR for two different elementary particles at relative rest (1/3)}

Consider two different elementary particles in flat spacetime.
We assume that both particles are at rest, or moving with the same velocity, in which
case we bring them to rest by performing a Lorentz transformation.

The two interacting point particles must be connected by a spacetime segment of null length (lightlike),
but for the time being we will not impose this condition. We start this derivation of the PAR
with a connecting spacetime segment of positive length (spacelike), and only in the
very last step of the proof we will go to the limit of a null length.
We are allowed to do this because the actual interaction takes place not between
points in Minkowski space, but between corresponding infinitesimal worldline segments
\cite{GaleriuArxiv2024}.
We notice that, after we go to the limit in the very last step, 
any two corresponding 
segments on the parallel worldlines of the two particles at rest
will have the same 
length, which means that $d\tau_1 = d\tau_2$, and the PAR simplifies to 
$\overrightarrow{\bf F_2} = - \overrightarrow{\bf F_1}$.

Let particle $A$ have the position four-vector $ (x_A, y_A, z_A, i\,c\,t_A)$
and let particle $B$ have the position four-vector $(x_B, y_B, z_B, i\,c\,t_B)$.
We assume that $t_A < t_B$. Without loss of generality we also assume that 
the rest mass $m_A$ of particle $A$ is greater than the rest mass $m_B$
of particle $B$, $m_A > m_B$.

Let $O$ be a point on line $AB$, on the outside of segment $AB$
and to the left (in the past),
such that 
\begin{equation}
m_A \times OA = m_B \times OB. \label{eq:condition4O}
\end{equation}
We translate the origin of our 
4D reference frame to point $O$. As a consequence, we are now able to write
\begin{eqnarray}
m_A \, x_A = m_B \, x_B, \label{eq:mAxAmBxB} \\
m_A \, y_A = m_B \, y_B, \\
m_A \, z_A = m_B \, z_B, \\
m_A \, i\,c\,t_A = m_B \, i\,c\,t_B.
\end{eqnarray}

By first squaring and then adding the above equations we obtain
\begin{equation}
m_A^2 \, ( x_A^2 + y_A^2 + z_A^2 - c^2 \, t_A^2 ) 
= m_B^2 \, ( x_B^2 + y_B^2 + z_B^2 - c^2 \, t_B^2 ),
\label{eq:condition4Osquared}
\end{equation}
which is the same thing as the square of equation (\ref{eq:condition4O}).

We calculate the radius $L$ of the hypersphere according to the formula
\begin{equation}
L^4 = ( x_A^2 + y_A^2 + z_A^2 - c^2 \, t_A^2 ) ( x_B^2 + y_B^2 + z_B^2 - c^2 \, t_B^2 ).
\label{eq:L4hypersphere}
\end{equation}
The radius of the hypersphere being the geometric mean of $OA$ and $OB$,
the hypersphere will intersect line $AB$ somewhere between points $A$ and $B$.

Based on equations (\ref{eq:condition4Osquared}) and (\ref{eq:L4hypersphere}), we conclude that
\begin{equation}
\frac{m_A}{m_B}
= \frac{L^2}{x_A^2 + y_A^2 + z_A^2 - c^2 \, t_A^2}
= \frac{x_B^2 + y_B^2 + z_B^2 - c^2 \, t_B^2}{L^2}.
\label{eq:mAmB}
\end{equation}

We perform a conformal inversion, according to formula (\ref{eq:4}).

As a result of the conformal inversion, 
particle $A$ acquires the initial coordinates of particle $B$
\begin{eqnarray}
x'_A = \frac{L^2}{x_A^2 + y_A^2 + z_A^2 - c^2 \, t_A^2} x_A = \frac{m_A}{m_B} x_A = x_B, \label{eq:xA} \\
y'_A = \frac{L^2}{x_A^2 + y_A^2 + z_A^2 - c^2 \, t_A^2} y_A = \frac{m_A}{m_B} y_A = y_B, \label{eq:yA} \\
z'_A = \frac{L^2}{x_A^2 + y_A^2 + z_A^2 - c^2 \, t_A^2} z_A = \frac{m_A}{m_B} z_A = z_B, \label{eq:zA} \\
i\,c\,t'_A = \frac{L^2}{x_A^2 + y_A^2 + z_A^2 - c^2 \, t_A^2} i\,c\,t_A 
= \frac{m_A}{m_B} i\,c\,t_A = i\,c\,t_B, \label{eq:tA} 
\end{eqnarray}
and particle $B$ acquires the initial coordinates of particle $A$
\begin{eqnarray}
x'_B = \frac{L^2}{x_B^2 + y_B^2 + z_B^2 - c^2 \, t_B^2} x_B = \frac{m_B}{m_A} x_B = x_A, \label{eq:xB} \\
y'_B = \frac{L^2}{x_B^2 + y_B^2 + z_B^2 - c^2 \, t_B^2} y_B = \frac{m_B}{m_A} y_B = y_A, \label{eq:yB} \\
z'_B = \frac{L^2}{x_B^2 + y_B^2 + z_B^2 - c^2 \, t_B^2} z_B = \frac{m_B}{m_A} z_B = z_A, \label{eq:zB} \\
i\,c\,t'_B = \frac{L^2}{x_B^2 + y_B^2 + z_B^2 - c^2 \, t_B^2} i\,c\,t_B 
= \frac{m_B}{m_A} i\,c\,t_B = i\,c\,t_A. \label{eq:tB} 
\end{eqnarray}

The Jacobian determinant at the initial position of particle A is
\begin{equation}
W_A = - \left( \frac{x_A^2 + y_A^2 + z_A^2 - c^2 t_A^2}{L^2} \right) ^4 = - \left( \frac{m_B}{m_A} \right) ^4,
\label{eq:JacobianA}
\end{equation}
and at the initial position of particle and B is
\begin{equation}
W_B = - \left( \frac{x_B^2 + y_B^2 + z_B^2 - c^2 t_B^2}{L^2} \right) ^4 = - \left( \frac{m_A}{m_B} \right) ^4.
\label{eq:JacobianB}
\end{equation}

Under conformal transformations the rest masses transform as a scalar density of weight 1/4.
As a result of the conformal inversion, 
particle $A$ acquires the initial rest mass of particle $B$
\begin{equation}
m'_A = |W_A|^\frac{1}{4} \, m_A = \frac{m_B}{m_A} \, m_A = m_B,
\label{eq:massA}
\end{equation}
and particle $B$ acquires the initial rest mass of particle $A$
\begin{equation}
m'_B = |W_B|^\frac{1}{4} \, m_B = \frac{m_A}{m_B} \, m_B = m_A,
\label{eq:massB}
\end{equation}
where the absolute value bars around the Jacobian determinants
are the result of another coordinate transformation
with negative Jacobian determinant, 
soon to follow the conformal inversion discussed here.

Under conformal transformations the covariant components of the four-velocity transform
as a vector density of weight $- 1/4$
\begin{equation}
U'_\nu = |W|^{-1/4} \, \frac{\partial x^\beta}{\partial x'^\nu} U_\beta,
\end{equation}
and, with the help of (\ref{eq:Jacobian}) and (\ref{eq:dxbdxpn}), and also (\ref{eq:xA})-(\ref{eq:tA}), we can write
\begin{eqnarray}
U'_{A \nu} 
= \frac{{r'_A}^2}{L^2} 
\left( \frac{L^2}{{r'_A}^2} \delta^\beta_{\ \nu} - \frac{2 L^2}{( {r'_A}^2 )^2} {x'_A}^\beta \, x'_{A \nu} \right) 
U_{A \beta}
= - \left( \frac{2}{r_B^2} x_B^\beta \, x_{B \nu} - \delta^\beta_{\ \nu} \right) U_{A \beta}, \\
U'_{B \nu} 
= \frac{{r'_B}^2}{L^2} 
\left( \frac{L^2}{{r'_B}^2} \delta^\beta_{\ \nu} - \frac{2 L^2}{( {r'_B}^2 )^2} {x'_B}^\beta \, x'_{A \nu} \right) 
U_{B \beta}
= - \left( \frac{2}{r_A^2} x_A^\beta \, x_{A \nu} - \delta^\beta_{\ \nu} \right) U_{B \beta},
\end{eqnarray}
where, in order to prevent any confusion, we have introduced the notation 
$x^\alpha x_\alpha = x^2 + y^2 + z^2 - c^2 t^2 \equiv r^2$.

Under conformal transformations the contravariant components of the four-velocity transform
as a vector density of weight 1/4
\begin{equation}
U'^{\mu} = |W|^{1/4} \, \frac{\partial x'^\mu}{\partial x^\alpha} U^{\alpha},
\end{equation}
and, with the help of (\ref{eq:Jacobian}) and (\ref{eq:dxpmdxa}), we can write
\begin{eqnarray}
{U'_A}^{\mu} 
= \frac{r_A^2}{L^2} 
\left( \frac{L^2}{r_A^2} \delta^\mu_{\ \alpha} - \frac{2 L^2}{( r_A^2 )^2} x_A^\mu \, x_{A \alpha} \right)
U_A^{\alpha}
= - \left( \frac{2}{r_A^2} x_A^\mu \, x_{A \alpha} - \delta^\mu_{\ \alpha} \right) U_A^{\alpha}, \\
{U'_B}^{\mu} 
= \frac{r_B^2}{L^2} 
\left( \frac{L^2}{r_B^2} \delta^\mu_{\ \alpha} - \frac{2 L^2}{( r_B^2 )^2} x_B^\mu \, x_{B \alpha} \right)
U_B^{\alpha}
= - \left( \frac{2}{r_B^2} x_B^\mu \, x_{B \alpha} - \delta^\mu_{\ \alpha} \right) U_B^{\alpha}.
\end{eqnarray}

Under conformal transformations the covariant components of the four-force transform
as a vector density of weight 1/4
\begin{equation}
F'_\nu = |W|^{1/4} \, \frac{\partial x^\beta}{\partial x'^\nu} F_\beta,
\end{equation}
and, with the help of (\ref{eq:Jacobian}) and (\ref{eq:dxbdxpn}), and also (\ref{eq:xA})-(\ref{eq:tA}), we can write
\begin{eqnarray}
F'_{A \nu} 
= \frac{L^2}{{r'_A}^2} 
\left( \frac{L^2}{{r'_A}^2} \delta^\beta_{\ \nu} - \frac{2 L^2}{( {r'_A}^2 )^2} {x'_A}^\beta \, x'_{A \nu} \right) 
F_{A \beta}
= - \frac{L^4}{r_B^4} \left( \frac{2}{r_B^2} x_B^\beta \, x_{B \nu} - \delta^\beta_{\ \nu} \right) F_{A \beta}, \\
F'_{B \nu} 
= \frac{L^2}{{r'_B}^2} 
\left( \frac{L^2}{{r'_B}^2} \delta^\beta_{\ \nu} - \frac{2 L^2}{( {r'_B}^2 )^2} {x'_B}^\beta \, x'_{B \nu} \right) 
F_{B \beta}
= - \frac{L^4}{r_A^4} \left( \frac{2}{r_A^2} x_A^\beta \, x_{A \nu} - \delta^\beta_{\ \nu} \right) F_{B \beta}.
\end{eqnarray}

Under conformal transformations the contravariant components of the four-force transform
as a vector density of weight 3/4
\begin{equation}
F'^{\mu} = |W|^{3/4} \, \frac{\partial x'^\mu}{\partial x^\alpha} F^{\alpha},
\end{equation}
and, with the help of (\ref{eq:Jacobian}) and (\ref{eq:dxpmdxa}), we can write
\begin{eqnarray}
{F'_A}^{\mu} 
= \frac{r_A^6}{L^6} 
\left( \frac{L^2}{r_A^2} \delta^\mu_{\ \alpha} - \frac{2 L^2}{( r_A^2 )^2} x_A^\mu \, x_{A \alpha} \right)
F_A^{\alpha}
= - \frac{r_A^4}{L^4} \left( \frac{2}{r_A^2} x_A^\mu \, x_{A \alpha} - \delta^\mu_{\ \alpha} \right) F_A^{\alpha}, \\
{F'_B}^{\mu} 
= \frac{r_B^6}{L^6} 
\left( \frac{L^2}{r_B^2} \delta^\mu_{\ \alpha} - \frac{2 L^2}{( r_B^2 )^2} x_B^\mu \, x_{B \alpha} \right)
F_B^{\alpha}
= - \frac{r_B^4}{L^4} \left( \frac{2}{r_B^2} x_B^\mu \, x_{B \alpha} - \delta^\mu_{\ \alpha} \right) F_B^{\alpha}.
\end{eqnarray}

We now notice that, since points $A$, $B$, and the origin $O$ are collinear, 
according to (\ref{eq:mAxAmBxB})-(\ref{eq:condition4Osquared}) it follows that
\begin{equation}
\frac{x_A^\beta \, x_{A \nu}}{r_A^2} 
= \frac{x_B^\beta \, x_{B \nu}}{r_B^2} 
= \frac{x_C^\beta \, x_{C \nu}}{r_C^2},
\end{equation}
where $x_C^\beta$ are the coordinates of any given point $C$ on the straight line $AB$,
placed on the same side of the origin $O$ as points $A$ and $B$. The ratios $x_C^\beta / r_C$
represent the direction cosines of ray $\overrightarrow{OC}$.

This confirms that the covariant and contravariant components of the transformed four-velocity are equal,
as they should, since the metric tensor $\eta_{\mu \nu}$ in Minkowski space 
$(x, y, z, i\,c\,t)$ is the identity matrix. The same conclusion applies to the 
covariant and contravariant components of the transformed four-force, 
since in this case we can also use equation (\ref{eq:L4hypersphere}).

To summarize, as a result of the conformal inversion discussed here, the contravariant components 
of the four-velocity and of the four-force become
\begin{eqnarray}
{U'_A}^{\mu} 
= - \left( \frac{2}{r_C^2} x_C^\mu \, x_{C \alpha} - \delta^\mu_{\ \alpha} \right) U_A^{\alpha}, \\
{U'_B}^{\mu} 
= - \left( \frac{2}{r_C^2} x_C^\mu \, x_{C \alpha} - \delta^\mu_{\ \alpha} \right) U_B^{\alpha}, \\
{F'_A}^{\mu} 
= - \frac{r_A^2}{r_B^2} \left( \frac{2}{r_C^2} x_C^\mu \, x_{C \alpha} - \delta^\mu_{\ \alpha} \right) F_A^{\alpha}, \\
{F'_B}^{\mu} 
= - \frac{r_B^2}{r_A^2} \left( \frac{2}{r_C^2} x_C^\mu \, x_{C \alpha} - \delta^\mu_{\ \alpha} \right) F_B^{\alpha}.
\end{eqnarray}

I am indebted to my wife for identifying the first term
inside the round brackets as the matrix of a projection operator, 
and for assisting me with some of the calculations
presented in the next section. 
Thus it became clear that the whole expression inside the round brackets
represents a reflection across the $AB$ axis, a coordinate transformation that, up to a factor of $- 1$,
has been named \lq\lq a-reflection\rq\rq\ by Nicholas Wheeler \cite{NicholasWheeler}.

\section{A review of reflection across an axis formulas}

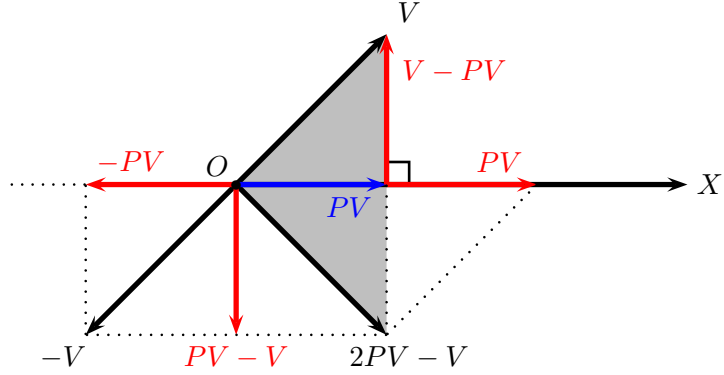
\begin{figure}[h!]
\begin{center}
\begin{pspicture}(-3,-3)(7,3)
\pspolygon*[linestyle=none,linecolor=lightgray](0,0)(2,0)(2,2)
\pspolygon*[linestyle=none,linecolor=lightgray](0,0)(2,0)(2,-2)
\psline[linewidth=2pt]{->}(0,0)(2,2)
\rput(2.3,2.3){$V$}
\psline[linewidth=2pt]{->}(0,0)(-2,-2)
\rput(-2.3,-2.3){$- V$}
\psline[linewidth=2pt]{->}(0,0)(2,-2)
\rput(2.3,-2.3){$2 P V - V$}
\psline[linewidth=2pt]{->}(0,0)(6,0)
\rput(6.3,0){$X$}
\psline[linewidth=1pt](2,0.3)(2.3,0.3)(2.3,0)
\psline[linewidth=2pt,linecolor=blue]{->}(0,0)(2,0)
\rput(1.5,-0.3){\color{blue}$P V$}
\psline[linewidth=2pt,linecolor=red]{->}(2,0)(4,0)
\rput(3.5,0.3){\color{red}$P V$}
\psline[linewidth=2pt,linecolor=red]{->}(2,0)(2,2)
\rput(2.9,1.5){\color{red}$V - P V$}
\psline[linewidth=2pt,linecolor=red]{->}(0,0)(-2,0)
\rput(-1.4,0.3){\color{red}$- P V$}
\psline[linewidth=2pt,linecolor=red]{->}(0,0)(0,-2)
\rput(0,-2.3){\color{red}$P V - V$}
\psline[linestyle=dotted,linewidth=1pt](-3,0)(-2,0)
\psline[linestyle=dotted,linewidth=1pt](-2,0)(-2,-2)
\psline[linestyle=dotted,linewidth=1pt](2,0)(2,-2)
\psline[linestyle=dotted,linewidth=1pt](-2,-2)(2,-2)
\psline[linestyle=dotted,linewidth=1pt](2,-2)(4,0)
\psdot(0,0)
\rput(-0.25,0.25){$O$}
\end{pspicture}
\caption{The reflection of vector $V$ across the axis of vector $X$.}
\label{fig:2}
\end{center}
\end{figure}

Consider a 4D Euclidean or pseudo-Euclidean $(x , y , z , i\,c\,t)$ vector space, and a position four-vector with components
$X^T = ( x_1 , x_2 , x_3 , x_4 )$. 
Due to the Euclidean metric tensor, no distinction is made in this section 
between covariant and contravariant components.
The square of the magnitude of this four-vector is 
$r^2 = X^T X = x_1^2 + x_2^2 + x_3^2 + x_4^2 = x_k x_k$, 
where the row $X^T$ is the transpose of the column $X$.
Assuming that $r^2 \neq 0$, we build the square matrix $P$ with elements $p_{i j} = x_i \, x_j / r^2$. Since
\begin{equation}
p_{i j} \, p_{j k} = \frac{x_i \, x_j}{r^2} \frac{x_j \, x_k}{r^2} 
= \frac{x_i \, x_j \, x_j\,  x_k}{r^4} = \frac{x_i \, r^2 \, x_k}{r^4} = \frac{x_i \, x_k}{r^2} = p_{i k},
\end{equation}
we conclude that matrix $P$ is idempotent, $P^2 = P$. 
The eigenvalues of an idempotent matrix are either 0 or 1. 
Indeed, if $P Y = \lambda Y$ then $P^2 \, Y = \lambda^2 \, Y$, 
and the eigenvalues $\lambda$ are solutions of the equation $\lambda^2 = \lambda$.
The sum of all the eigenvalues
is equal to the trace of the matrix. In our case
\begin{equation}
\textrm{Tr} \, P = p_{i i} = \frac{x_i \, x_i}{r^2} = \frac{r^2}{r^2} = 1.
\end{equation}
We conclude that matrix $P$ has only one eigenvalue equal to 1 and three eigenvalues equal to 0.
A matrix that is idempotent and symmetric is a projection matrix. 
In our case, matrix $P$ projects any vector $V$ along the direction of vector $X$.
This can be most clearly seen from the fact that $X$ is the eigenvector corresponding to the eigenvalue of 1. Indeed, $P X = X$, as can be checked by direct calculation 
\begin{equation}
p_{i j} \, x_j = \frac{x_i \, x_j}{r^2} x_j = \frac{x_i \, x_j \, x_j}{r^2} = \frac{x_i \, r^2}{r^2} = x_i.
\end{equation}
Next we build the square matrix $R = 2 P - I$, where $I$ is the identity matrix. Matrix $R$ has the same eigenvectors as matrix $P$. Indeed,
if $P Y = \lambda Y$, then $R Y = 2 P Y - I Y = 2 \lambda Y - Y = ( 2 \lambda - 1 ) Y$. For an eigenvector parallel to $X$, the eigenvalue of $P$ is 1,
and the eigenvalue of $R$ is also 1. For an eigenvector perpendicular to $X$, the eigenvalue of $P$ is 0, and the eigenvalue of $R$ is $-1$.
The determinant of $R$ is equal to the product of the eigenvalues of $R$, which in our case is $-1$. Matrix $R$ is symmetric, $R^T = R$, and we also 
notice that $R^2 = I$, as can be checked by direct calculation
\begin{multline}
r_{i j} \, r_{j k} = \left( 2 \frac{x_i \, x_j}{r^2} - \delta_{i j} \right) \left( 2 \frac{x_j \, x_k}{r^2} - \delta_{j k} \right) \\
= 4 \frac{x_i \, x_j \, x_j \, x_k}{r^4} -  2 \frac{x_i \, x_j}{r^2} \delta_{j k} - 2 \frac{x_j \, x_k}{r^2} \delta_{i j} + \delta_{i j} \, \delta_{j k} \\
= 4 \frac{x_i \, r^2 \, x_k}{r^4} - 2 \frac{x_i \, x_k}{r^2} - 2 \frac{x_i \, x_k}{r^2} + \delta_{i k} = \delta_{i k},
\end{multline}
where $\delta_{i k}$ is the Kronecker delta.
We conclude that matrix $R$ is an improper rotation matrix, since $R^T = R^{-1}$ and the determinant is $-1$.
This improper rotation is in fact a reflection across the axis of the position four-vector $X$, as can be seen in Figure \ref{fig:2}, 
where the position four-vector $V$ is transformed into $R V = 2 P V - V$. 
This 4D transformation is very similar to the 3D Euclidean analogue,
where a reflection across an axis is a proper rotation of $180^{\circ}$ around that axis.
Since $R$ is a rotation matrix, it follows that $|V|$, the length of $V$, is unchanged by this transformation. 
This can also be seen in Figure \ref{fig:2}, 
where we have two congruent right triangles
with hypotenuses of length $|V|$ and $|2 P V - V|$,
and with perpendicular sides of length $|PV|$ and $|V - P V|$.

We now recognize the fact that the parity transformation $(x \to - x , y \to - y , z \to - z)$ is a special case of this
kind of improper rotation in 4D, it is a reflection across the time axis. But since in Special Relativity there is nothing special about the
time axis of a given reference frame, it follows that the parity symmetry in fact extends to the more general case of reflections across any axis.

For completeness we should also mention that, due to the transitivity of the Jacobian,  two such
reflections across an axis (improper rotations) in 4D
make up a proper rotation, that is a proper Lorentz transformation \cite{EGPeterRowe2001}.

\section{A derivation of the PAR for two different elementary particles at relative rest (2/3)}

Now we return to the spacetime configuration that has resulted from the conformal inversion of Section 4,
and we perform a reflection across the $AB$ axis, according to
\begin{equation}
x''^\sigma = \left( \frac{2}{r_C^2} x_C^\sigma x_{C \mu} - \delta^\sigma_{\ \mu} \right) x'^\mu.
\label{eq:99}
\end{equation} 
The inverse transformation looks just like the original transformation, 
the reflection across an axis operation being the inverse of itself
\begin{equation}
x'^\mu = \left( \frac{2}{r_C^2} x_C^\mu x_{C \rho} - \delta^\mu_{\ \rho} \right) x''^\rho.
\label{eq:100}
\end{equation} 
The coordinates of point $C$ are left unchanged by 
either (\ref{eq:99}) or (\ref{eq:100}), and for this reason we keep them
without prime or double prime notation.
The coordinates of particles $A$ and $B$ are also left unchanged.

This is a linear and homogeneous coordinate transformation, just like the well known 
Lorentz transformation from Special Relativity, but improper.
The Jacobian matrix, whose elements appear inside the round brackets,
has a determinant of $- 1$, as calculated in Section 5, 
just as needed.

As a result of the reflection across the $AB$ axis, the contravariant components 
of the four-velocity and of the four-force become
\begin{multline}
{U''_A}^{\sigma} 
= \left( \frac{2}{r_C^2} x_C^\sigma x_{C \mu} - \delta^\sigma_{\ \mu} \right) {U'_A}^\mu \\
= - \left( \frac{2}{r_C^2} x_C^\sigma x_{C \mu} - \delta^\sigma_{\ \mu} \right)
\left( \frac{2}{r_C^2} x_C^\mu x_{C \alpha} - \delta^\mu_{\ \alpha} \right) U_A^{\alpha}, \\
= - \delta^\sigma_{\ \alpha} U_A^{\alpha} = - U_A^{\sigma},
\end{multline}
\begin{eqnarray}
{U''_B}^{\sigma} 
= \left( \frac{2}{r_C^2} x_C^\sigma x_{C \mu} - \delta^\sigma_{\ \mu} \right) {U'_B}^\mu
= - U_B^{\sigma}, \\
{F''_A}^{\sigma} 
= \left( \frac{2}{r_C^2} x_C^\sigma x_{C \mu} - \delta^\sigma_{\ \mu} \right) {F'_A}^\mu
= - \frac{r_A^2}{r_B^2} F_A^{\sigma}, \label{eq:5_103} \\
{F''_B}^{\sigma} 
= \left( \frac{2}{r_C^2} x_C^\sigma x_{C \mu} - \delta^\sigma_{\ \mu} \right) {F'_B}^\mu
= - \frac{r_B^2}{r_A^2} F_B^{\sigma}. \label{eq:5_104}
\end{eqnarray}

Now the spacetime configuration very much resembles that from Section 2.
The two elementary particles have swapped their position and time coordinates, and also their rest masses,
while at the same time keeping the same common tangent to their worldlines.
As a consequence, due to symmetry, and also due to our additional assumption that the four-forces
do not depend on four-accelerations or higher derivatives, the following two relations must hold
\begin{eqnarray}
{F''_A}^\sigma = F_B^\sigma, \label{eq:5_105} \\
{F''_B}^\sigma = F_A^\sigma. \label{eq:5_106}
\end{eqnarray}

Due to equations (\ref{eq:5_103})-(\ref{eq:5_104}), 
either of equations (\ref{eq:5_105})-(\ref{eq:5_106}) is equivalent to
\begin{equation}
\overrightarrow{\bf F_B} = - \frac{r_A^2}{r_B^2} \overrightarrow{\bf F_A}.
\label{eq:5_107}
\end{equation}

This result may seem a little bit disappointing at first, due to the unexpected fraction,
until we realize that in the real world the interaction happens between particles separated by a null spacetime interval,
while the whole demonstration so far has assumed a positive spacetime interval.
How do we go to the limit of a null interval, when $r_A \to 0$ and $r_B \to 0$,
and when the fraction $r_A / r_B$ could very well converge to 1?
Our optimism is justified by the fact that this is not the first time when, 
due to the indefinite metric of Minkowski space, we have encountered such 
a fraction of two geometrical measures that both converge to zero in the case of practical importance.
Nonetheless, by evaluating what happens when we go to the limit, we were able to derive the 
Doppler factor in the Li\'{e}nard-Wiechert potentials \cite{Galeriu2021EJP, Galeriu2021Arxiv}.

\section{A derivation of the PAR for two different elementary particles at relative rest (3/3)}

Without loss of generality we can assume that the two particles at rest have coordinates $(x_A, 0, 0, i\,c\,t_A)$ and $(x_B, 0, 0, i\,c\,t_B)$, 
otherwise we can always rotate the 3D reference frame in order to make it happen.

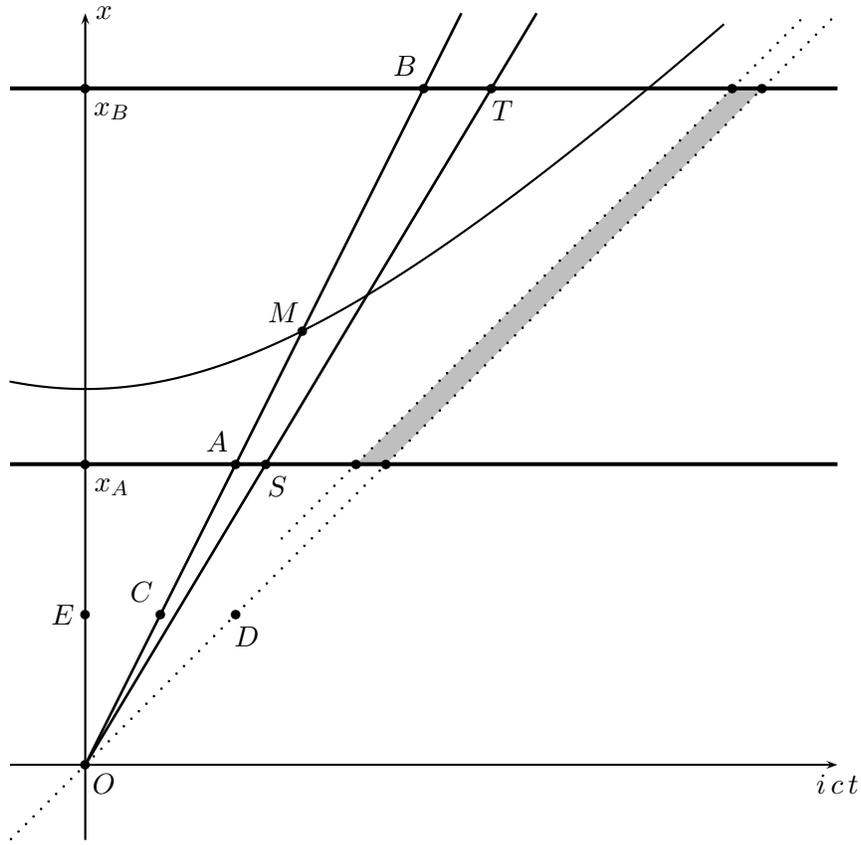
\begin{figure}[h!]
\begin{center}
\begin{pspicture}(-1,-1)(10,10)
\pspolygon*[linestyle=none,linecolor=lightgray](4,4)(9,9)(8.6,9)(3.6,4)
\psdot(0,0)
\rput(0.25,-0.25){$O$}
\psline{->}(-1,0)(10,0)
\rput(10,-0.25){$i\,c\,t$}
\psline{->}(0,-1)(0,10)
\rput(0.25,10){$x$}
\psline[linestyle=dotted,linewidth=1pt](-1,-1)(10,10)
\psdot(0,2)
\rput(-0.3,2){$E$}
\psdot(2,2)
\rput(2.15,1.7){$D$}
\psline[linewidth=1.5pt](-1,4)(10,4)
\psdot(0,4)
\rput(0.35,3.7){$x_A$}
\psline[linewidth=1.5pt](-1,9)(10,9)
\psdot(0,9)
\rput(0.35,8.7){$x_B$}
\psparametricplot[plotstyle=curve]{-0.2}{1.3}{t SINH 5 mul t COSH 5 mul}
\psline[linewidth=1pt](0,0)(5,10)
\psdot(2,4)
\rput(1.75,4.3){$A$}
\psdot(4.5,9)
\rput(4.25,9.3){$B$}
\psdot(2.88675,5.77350)
\rput(2.63675,6.02350){$M$}
\psdot(1,2)
\rput(0.75,2.3){$C$}
\psline[linewidth=1pt](0,0)(6,10)
\psdot(2.4,4)
\rput(2.55,3.7){$S$}
\psdot(5.4,9)
\rput(5.55,8.7){$T$}
\psline[linestyle=dotted,linewidth=1pt](2.6,3)(9.6,10)
\psdot(3.6,4)
\psdot(4,4)
\psdot(8.6,9)
\psdot(9,9)
\end{pspicture}
\caption{In the limit, when the interacting particles $AS$ and $BT$
become two corresponding segments of infinitesimal length,
the two parallel worldlines of the two particles at rest,
together with the two parallel null lines connecting the endpoints of the two segments,
will make an infinitely thin parallelogram.}
\label{fig:3}
\end{center}
\end{figure}

Through the origin $O$ we draw another line very close to line $AB$. This line intersects the 
worldline of the first particle in point $S$, and the worldline of the second particle in point $T$, as seen in Figure \ref{fig:3}.
This construction is needed because the actual electrodynamic interaction happens not between points in Minkowski space,
but between corresponding length elements along the worldlines of the two particles \cite{GaleriuArxiv2024}.
Our interacting point particles $AS$ and $BT$, although with a null spatial extent, have a non-null temporal extent.
In the limit, when endpoints $A$ and $B$, and also endpoints $S$ and $T$, are connected by null lines (by light signals),
the two worldline segments $AS$ and $BT$ will become corresponding segments of infinitesimal length.

Since the two particles are at rest and their worldlines are parallel, we notice that we have two
similar triangles,  $\Delta OAS \sim \Delta OBT$, and as a result
\begin{equation}
\frac{r_A}{r_B} = \frac{OA}{OB} = \frac{OS}{OT} = \frac{AS}{BT}.
\label{eq:6_108}
\end{equation}

With the help of (\ref{eq:6_108}) we turn equation (\ref{eq:5_107}) into
\begin{equation}
\overrightarrow{\bf F_B} = - \left(\frac{AS}{BT}\right)^2 \overrightarrow{\bf F_A},
\label{eq:6_109}
\end{equation}
and we turn equation (\ref{eq:condition4O}) into
\begin{equation}
m_A \times OS = m_B \times OT, \label{eq:condition4Onew}
\end{equation}
which means that, if instead of starting with the two interacting point particles 
at $A$ and $B$ we start with them at $S$ and $T$,
we end up with the same location for the origin $O$. 
The same reference frame is being used throughout the whole process of
going to the limit.

When we go to the limit, 
not only the spacetime interval $AB$ will become null,
but also the spacetime interval $ST$.
In the limit the two lines $AB$ and $ST$ will become parallel to each other, having both the direction of line $OD$,
where $D$ is a given point on the bisecting line of the first quadrant, as seen in Figure \ref{fig:3}. 
How could these two parallel lines intersect at point $O$, the reader may ask,
shouldn't two parallel lines intersect at infinity? Well, two parallel lines intersect at infinity in the Euclidean plane,
but here we are in the Minkowski plane. The distance between any two points on line $OD$ is the same, still zero,
no matter how far or how close the two points may seem to be on the Minkowski diagram.
The Minkowski diagram shown in Figure \ref{fig:3} is only a Euclidean representation of the actual Minkowski plane.
Another challenge to our intuition is brought by the fact that $\angle BOD$ and $\angle TOD$ are of infinite measure.
Indeed, for $y = 0$ and $z = 0$, the intersection of the hypersphere of radius $L$ with the $(xOict)$ Minkowski plane
results in a \lq\lq circle\rq\rq, which in the Euclidian representation of Figure \ref{fig:3} is 
the hyperbola through point $M$ \cite{Galeriu2003}.
Since line $OD$ is an asymptote of this hyperbola, and these two curves do not intersect at finite coordinates, 
the arclength along the hyperbola branch will increase without limit. 
In terms of angles the whole process of going to the limit
consists of three simultaneous limits, $\angle EOB \to \infty$, $\angle EOT \to \infty$, and $\angle BOT \to 0$,
where $E$ is a given point on the positive $x$ axis, as seen in Figure \ref{fig:3}.

Due to the infinitely thin parallelogram that appears in the limit, we find that $AS / BT \to 1$, which is the expected result.
In this way equation (\ref{eq:6_109}) reproduces the PAR for two particles at relative rest.

\section{ Concluding remarks}

The guiding idea throughout this research project was that, 
in a theory of time-symmetric action-at-a-distance interactions,
if we can find a coordinate transformation that can
swap the position and time coordinates of the two interacting particles, together with their rest masses,
electric charges (if different), four-velocities, four-accelerations, and higher derivatives (if needed),
then, due to symmetry considerations, a relationship between the advanced and retarded four-forces
should emerge, leading to the principle of action and reaction.
Although at first sight this mission might seem impossible, the problem is greatly simplified by the
supplementary hypothesis that the four-forces depend only on the rest masses, the electric charges,
the position four-vectors, and the four-velocities of the two interacting particles, hypothesis
supported by geometrical considerations in Minkowski space \cite{GaleriuArxiv2024}.
We have studied a coordinate transformation consisting of
a conformal inversion
followed by
a reflection across the axis connecting the two interacting point particles.
Such a coordinate transformation could also be expressed in terms of special conformal transformations,
translations, and improper a-reflections \cite{NicholasWheeler}.
Helped by the fact that the rest masses change under conformal transformations, 
we were able to only partially achieve our goal. 
Our coordinate transformation was not able to also swap the four-velocities of the two interacting particles.
This is due to the fact that a conformal transformation, by its name giving property, will preserve the angles 
between the four-velocities of each of the two interacting particles and the connecting spacetime segment.
In consequence, our demonstration applies only to particles at relative rest, either at rest or moving with
the same velocity.
Knowing that the principle of action and reaction holds even when the two four-velocities are different,
we make the conjecture that another, undetermined yet, symmetry operation must also apply.

\end{document}